\documentclass[aps,prl,twocolumn,superscriptaddress]{revtex4}
\usepackage{graphicx}

\usepackage{amsmath}

\usepackage[draft]{hyperref} 
\usepackage{bm,amsfonts}
\usepackage{graphicx}

\mathchardef\varPhi="0108
\begin{document}


\title{Fringe free holographic measurements of large amplitude vibrations}

\author{ F. Joud$^1$,  F. Verpillat$^1$, F. Lalo\"{e}$^1$,  M. Atlan$^2$, J. Hare$^1$ and M. Gross$^1$}

\address{ $^1$ Laboratoire Kastler-Brossel; UMR 8552 CNRS,
\'Ecole Normale Sup\'erieure, Universit\'e Paris 6; 24 rue Lhomond 75231 Paris Cedex 05 France }

\address{$^2$ Institut Langevin: UMR 7587 CNRS INSERM, ESPCI ParisTech, Universit\'e Paris 6, Universit\'e Paris 7, 10
rue Vauquelin, 75 231 Paris Cedex 05, France.}


\begin{abstract}
In the measurement of the amplitude  of vibration of objects,
holographic imaging techniques usually involve fringe counting;
due to the limited resolution of images, measurements of large
amplitudes are not accessible. We demonstrate a technique that
suppresses the necessity of fringe counting: frequency sideband
imaging, where the order of the sideband is considered as a marker
of the amplitude. The measurement is completely local: no
comparison with another reference point on the object is
necessary. It involves a sharp variation of a signal, which makes
it robust against perturbations. The method is demonstrated in an
experiment made with a vibrating clarinet reed; phase modulations
as large as $1000$ radians have been measured.
\end{abstract}

%


\maketitle

ocis{090.1760,200.4880,040.2840,100.2000}


The observation of interference fringes in holographic methods provides accurate measurements of the amplitude of
vibration of objects. Powell and  Stetson \cite{powell1965iva} have shown that the fringes on the holographic
reconstruction of a vibrating object correspond, after time averaging, to zeros of the Bessel function
$J_{0}(\varPhi)$, where $\varPhi(x,y)$ is the amplitude of phase modulation of the optical field emitted by the object
at point $x,y$. Digital holography was introduced in 1994 by Schnars and J\"{u}pner \cite{Schnars_Juptner_94}. In 2003
Picard et al. \cite{picart2003tad} transposed the time averaging method to digital holography.

In a previous letter \cite{gross_anche_2008}, we described sideband digital holography, based on the detection of the
light back-scattered by a vibrating object at different sideband frequencies;  the fringes for sideband $n$ then
correspond to the zeros of the $n$-th order Bessel function $J_{n}(\varPhi)$. As in the work made by Aleksoff
\cite{aleksoff1971tuh} in 1971, the reference beam was frequency shifted to select one sideband $n$, but the use of
acousto-optic modulators and numerical techniques provided much more flexibility. In reference~\cite{gross_anche_2008},
we have shown how the comparison of dark fringes for different sideband leads to a determination of the vibration
amplitude $\varPhi(x,y)$ at each point of the object. This determination is non-local, since  it involves counting
fringes from one reference point of the image to the point of interest, so that large amplitudes are not accessible.

In this letter, we demonstrate another approach   that completely eliminates the necessity of counting fringes; it
gives a local measurement of the amplitude of vibration, even for large values. For any pixel of coordinate $x,y$, we
consider the sideband order $n$ as a variable and we  plot the intensity $I$ as a function of $n$. One can then easily
determine $\varPhi$ since $I(n)$ exhibits a sharp variation from maximum to zero near $n \simeq \varPhi$. The method is
robust and can easily be used even when the fringes become so narrow that they cannot be resolved, which gives
immediate access to large amplitudes of vibration.

Consider a point of the object vibrating  at frequency $\nu_A$ and amplitude $z_{\text{\tiny max}}$; its displacement
$z(t)$ is:
\begin{eqnarray}\label{Eq_1}
z(t) = z_{\text{\tiny max}}\sin(2 \pi \nu_A t )
\end{eqnarray}
In backscattering geometry, this corresponds to a phase
modulation:
\begin{eqnarray}\label{Eq_2}
  \varphi(t)=4 \pi z(t)/\lambda= \varPhi \sin(2 \pi \nu_A t )
\end{eqnarray}
where $\lambda$ is the optical wavelength and $\varPhi = 4 \pi z_{\text{\tiny max}}/\lambda$. The scattered field is then:
\begin{eqnarray}\label{Eq_6}
  E(t)= {\cal E} e^{j\left( \nu_0 t + \varphi(t)\right)} ={\cal E} \sum_n J_n(\varPhi ) e^{j (\nu_0+n \nu_A)t}
\end{eqnarray}
where ${\cal E}$ is the complex amplitude of the field, $\nu_0$ the frequency of the illumination optical field,
and $J_n$ the $n$-th order Bessel function of the first kind;
$J_{-n}(z)=(-1)^{n}\; J_{n}(z)$ for  integer $n$ and  real $z$.
The scattered field is then the sum of sidebands with frequencies
$\nu_0 + n \nu_A$ and intensities $ I_n$ given by:
\begin{eqnarray}\label{Eq_9}
       I_n(\varPhi)= \left|{\cal E}J_n(\varPhi)\right|^2
\end{eqnarray}

\begin{figure}[]
\centering
\includegraphics[width=8.5cm]{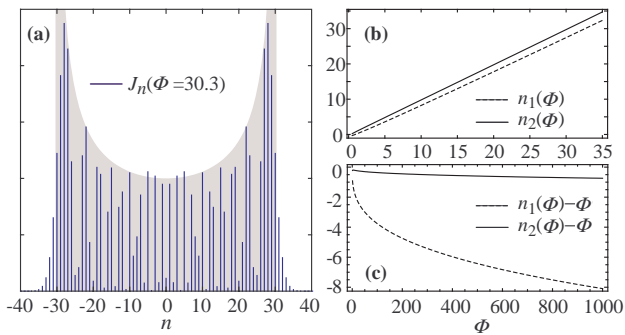}
\caption{(a) Relative intensities of the sidebands  as a function of $n$ for fixed $\varPhi=30.3$~rad. The vertical lines show
the intensities of the discrete $n$ components of the real spectrum. The light grey shade shows the Doppler spectrum
obtained from the vibration velocity distribution, with  a continuous variable on the horizontal axis $ n=(\nu-\nu_{0})
/\nu_{A}$.  Both spectra fall abruptly beyond $n= \pm 30.3$, which corresponds to the Doppler shift associated with the
maximum velocity. (b) The broken line shows the  values $n_1$ as a function of $\varPhi$, where $n_1$ is the value of
$n$ giving the maximum intensity in the discrete spectrum; the full line shows $n_2$, where $n_2$ is the value of $n$
for which the intensity if half the maximum, which gives a very good approximation of $\varPhi$. (c) The broken and
full lines respectively show $n_1-\varPhi$ and $n_2-\varPhi$ as a function of $\varPhi$. } \label{Fig1}
\end{figure}

Figure~\ref{Fig1}(a) shows the intensity  of the sidebands as a function of $n$, assuming
$\varPhi=30.3$~rad. When $n$ is considered as a continuous variable $n=( \nu-\nu_{0}) /\nu_{A}$ giving the Doppler
frequency shift in units of  $\nu_A$, one obtains the light grey shade corresponding to the Doppler spectrum deduced
from the vibration velocity distribution. This continuous spectrum is proportional to $ [\;1- (n/\varPhi)^2 ]^{-1/2}$,
where $n$ is confined between the values $\pm \varPhi$ that correspond to the maximum Doppler shift. The discrete
spectrum has a similar behavior, remaining mostly confined between the same $n$ values, and dropping abruptly from a
maximum reached close to $n = \pm \varPhi$ to almost zero. This is the key idea of our method: we measure the frequency
position of this sharp variation and deduce from it the value of the vibration amplitude. This method can be seen as a
discrete spectrum version of laser Doppler imaging of non-periodic motions reported in
\cite{atlan2006ldi,atlan2006fdw}.

Figs.~\ref{Fig1}(b,c) give more detail  on this sharp variation. For each value of $\varPhi$ we calculate the value
$n_1(\varPhi)$ of $n$ that, in Eq.~(\ref{Eq_9}), gives the maximum intensity. To avoid the quantization noise induced
by discrete variables, the calculation is made with Bessel functions of fractional order, but of course only integer
values of $n$ are relevant to the experiment. It is known \cite{abramowitz1965hmf} that $n_1(\varPhi) \simeq \varPhi$, within a correction of
order $\varPhi^{1/3}$ ; the broken lines shows $n_1(\varPhi)$ (b), and the difference
$n_1(\varPhi)-\varPhi$ (c). To determine more precisely the location of the abrupt drop from maximum to zero, we
calculate the $n$ value $n_2(\varPhi)$ for which the intensity is half the maximum. The $n_2(\varPhi)$ and the
difference $n_2(\varPhi)-\varPhi$ are shown by the full line in Fig.~\ref{Fig1} (b) and (c) respectively. Clearly,
$n_2$ gives an accurate evaluation of $\varPhi$.

The experimental setup is the same as described in Ref.~\cite{gross_anche_2008}. As in ~\cite{gross_anche_2008}, we
have chosen a clarinet reed as the vibrating object to experimentally demonstrate the method. The  reed, vibrating on
its first flexural resonance mode at frequency $\nu_A \sim 2100$  Hz, is illuminated at $\lambda =650$ nm by a laser
field $E_I$ at frequency $\nu_0$. The CCD camera (frame frequency $\nu_{CCD} =10 $ Hz) records the interference pattern
(the hologram) between the scattered light and the local oscillator beam (field $E_{LO}$, frequency $\nu_{LO}$). Two
acousto-optic modulators (Bragg cells) with frequencies $\nu_{AOM1}$ and $\nu_{AOM2}$ are used to adjust the
frequencies $\nu_0=\nu_L + \nu_{AOM2}$ and $\nu_{LO}=\nu_L + \nu_{AOM1}$; an arbitrary sideband $n$ can then be
selected by adjusting these frequencies. A 4-phase detection of sideband $n$ is obtained by adjusting them to fulfil
the relation:
\begin{equation}\label{Eq_10_}
  \nu_0+ n \nu_A -\nu_{LO}(n)= \nu_{CCD}/4
\end{equation}

We record  a sequence of consecutive CCD images $I_0,I_1,...I_{M-1}$, where $M$ is a multiple of 4.  The complex
hologram $H$ in the plane of the CCD is is then:

\begin{equation}\label{Eq_11_}
  H=\sum_{m=0}^{M-1} j^m~ I_m
\end{equation}

\begin{figure}[]
\centering
\includegraphics[width=8.5cm]{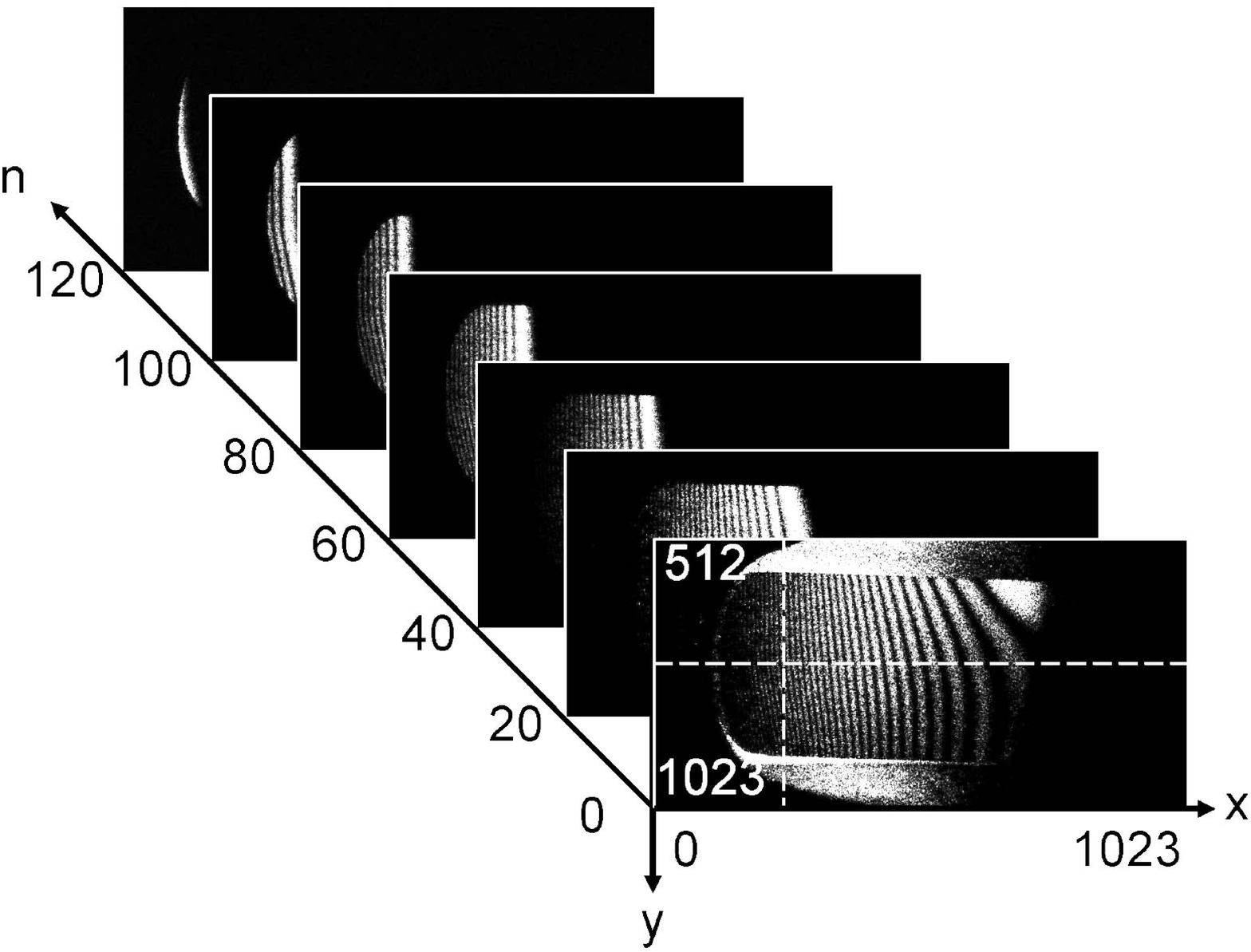}
\caption{Cube of data obtained  from the reconstructed holographic images of a vibrating clarinet reed; sideband images
with $n=0,20,40...120$ are shown in arbitrary linear scale. By choosing $n$, one moves the border of the illuminated
region on the object, obtaining a local marker of  the amplitude of vibration. The white dashed lines correspond to
$x=249$ and $y=750$, i.e. to the point chosen for Fig.~\ref{Fig_enveloppes}.} \label{fig4_manip}
\end{figure}

From  $H$, the images are reconstructed as in Ref. \cite{gross_anche_2008} by the standard convolution method
\cite{Schnars_Juptner_94}, which provides the map of the complex field $E(n,x,y)$ in the plane of the object. By
successively adjusting the frequency  $\nu_{LO}(n)$ of the local oscillator to appropriate values, we then record the
intensity images $|E(x,y,n)|^2$ of the sidebands as a function of $x,y$ and $n$. We then obtain a cube of data with
three axes $x,y$ and $n$,  where $x,y$ are expressed in units of pixels of the reconstructed image of the reed.
Figure~\ref{fig4_manip} shows the images  obtained for for $n=0, 20,...120$ that correspond to cuts of the cube along
$x,y$ planes. The right part of the reed  ($x>800)$ is clamped on the mouthpiece. The images illustrate how, when $n$
increases, the fringes move towards regions with larger amplitudes of vibrations: no signal is obtained in regions
where $\varPhi=4 \pi z_{\text{\tiny max}}/\lambda \le n$. This well known property of Bessel functions allows one to
get a marker on the object, signalling regions where the amplitude correspond to $\varPhi \simeq 120$~rad.
\begin{figure}[]
\centering
\includegraphics[width=8.5cm]{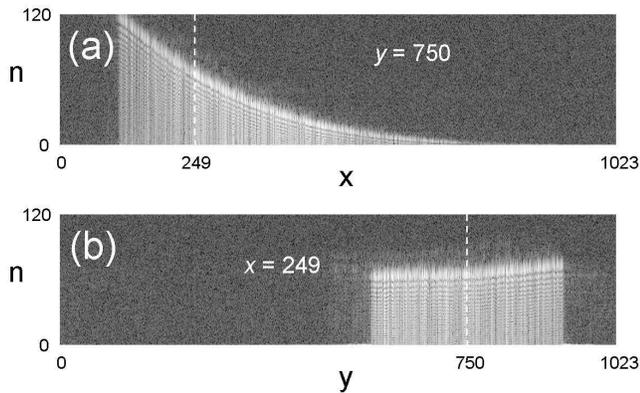}
\caption{Images corresponding to cuts of 3D data of the reconstructed images along the planes $y=750$ (a) and $x=249$
(b). Fig~(a) shows the deformation of the object along its axis and Fig~(b) a transverse cut with  a slight vibration
asymmetry. A logarithmic intensity scale is used.} \label{Fig_enveloppes}
\end{figure}
Fig.~\ref{Fig_enveloppes}(a) displays a cut of the cube of data along the $x,n$ horizontal plane $y=750$ (horizontal
white dashed line in  Fig.~\ref{fig4_manip}). The envelope of the non-zero (non black) part of the image provides a
measurement of the amplitude of vibration in units  of $\lambda/4 \pi$. We actually obtain a direct visualization of
the shape of the reed at maximal elongation, from the right part clamped on the mouthpiece to the tip on the left. The
maximum amplitude correspond to $\varPhi \simeq 120$~rad. Fig.~\ref{Fig_enveloppes}(b) displays a transverse cut along
the $y,n$ vertical  $x=249$ plane (vertical white  dashed line in Fig.~\ref{fig4_manip}); a slight asymmetry of the
reed vibration is clearly visible.

\begin{figure}[]
\centering
\includegraphics[width=8.0cm]{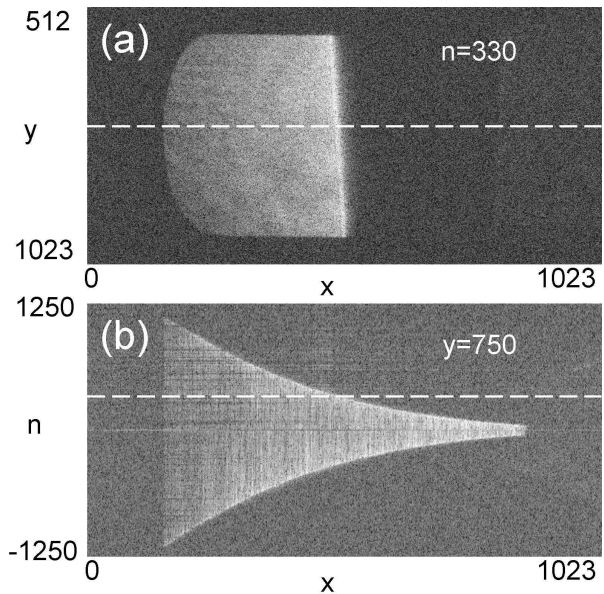}
\caption{(a): image reconstructed with sideband $n=330$, with a large amplitude of vibration. (b) is the equivalent of
Fig.~\ref{Fig_enveloppes}(a), but with positive  and negative $n$ values; one measures a maximum vibration amplitude of
$z_{\text{\tiny max}} \simeq 60 \: \mu$m. A logarithmic intensity scale is used.} \label{Fig_enveloppes_large}
\end{figure}

Figures~\ref{fig4_manip} and ~\ref{Fig_enveloppes} have been obtained by exciting the reed at the resonant frequency
$\nu_A=2123$~Hz and at moderate excitation level. The consistency of amplitude measurements obtained in this way with those obtained by fringe counting is illustrated in Fig. 3 of Ref. \cite{gross_anche_2008}. We have also used higher excitation amplitudes, about $10$ times
larger (the frequency of reed resonance is then shifted to  $\nu_A=2020$ Hz). Figure~\ref{Fig_enveloppes_large} (a)
shows the images obtained for $n=330$: the fringes are now completely unresolved, but the transition from zero to
non-zero intensity remains very clear. With a single hologram, and without fringe counting, one obtains a clear marker
of the line where $ \varPhi(x,y)=330$~rad.

Fig.~\ref{Fig_enveloppes_large}(b) shows the equivalent of Fig.~\ref{Fig_enveloppes}(a), but with a higher excitation
level, and this time for positive and negative values of $n$; data range up to about $|n| \simeq 1140$,
corresponding to $z_{\text{\tiny max}} \simeq 58.4 \: \mu$m.  Since the vibration amplitude is much larger than $\lambda$, the continuous approximation for $n$ is valid, and the images of
Fig. ~\ref{Fig_enveloppes} can be reinterpreted in term of classical Doppler effect: if the signal intensity $I_n(x,y)$ at frequency $\nu_0+n\nu_A$ is non-zero, at some time $t$ of the periodic motion the Doppler shift $\nu_0+2
v(x,y,t)/c$  is equal to  $n\nu_A$ ($v$ and $c$ are  the reed and light velocities). The accuracy of the measurement of the amplitude corresponds to about $0.5 \mu$ (see discussion in the supplementary material). A fine analysis of the 3D data shows  a slight asymmetry  between the positive and negative $n$ values. In
Fig.~\ref{Fig_enveloppes_large}(b) for example, the maximum value for $|n|$ is about 1160 for $n>0$, and 1120 for $n<0$.
This means that the motion of the reed in not perfectly sinusoidal, and that maximum velocities (i.e. maximum Doppler
shifts) are slightly different for the up and down motion of the  reed. A more detailed study of the modification of the spectrum due to the superposition of several harmonic motions can be done, but falls beyond the scope of this letter.

In conclusion, taking advantage of the  frequency (or the sideband order $n$) of the light scattered by a vibrating
object adds a new dimension to digital holography. Each pixel of the image then provides an information that is
completely independent of the others, which results in redundancy and robustness of the measurements. Looking at the
edges  of the spectrum provides an accurate determination of the vibration amplitude and avoids a cumbersome analysis
of the whole cube of data cube, giving easy access to a measurements of large amplitudes of oscillation. Our method
could be combined with other techniques, such as phase modulation of the reference beam
\cite{lokberg1994rse,zhang2004vap}, in order to get information on the phase of the mechanical vibration without
loosing the robustness of the measurement.

\end{document}